\documentclass[pra,twocolumn,amsmath,amssymb,floatfix,superscriptaddress]{revtex4-1}
\usepackage{graphicx}
\usepackage{dcolumn}
\usepackage{bm}
\usepackage{amsthm}
\usepackage{amsmath}
\usepackage{amssymb}
\usepackage{color}
\usepackage{float}
\usepackage{placeins}

\definecolor{Red}{rgb}{1,0,0}

\def\ket#1{| #1 \rangle}

\begin{document}

\title{Relative performance of ancilla verification and decoding in the [[7,1,3]] Steane code}

\author{Ali Abu-Nada}
\affiliation{Department of Physics, Southern Illinois University, Carbondale, IL 62901, USA}
\author{Ben Fortescue}
\affiliation{Department of Physics, Southern Illinois University, Carbondale, IL 62901, USA}
\author{Mark Byrd}
\affiliation{Department of Physics, Southern Illinois University, Carbondale, IL 62901, USA}
\affiliation{Department of Computer Science, Southern Illinois University, Carbondale, IL 62901, USA} 
\date{June 3, 2014}

\begin{abstract}
Ancilla post-selection is a common means of achieving fault-tolerance in quantum error-correction.  However, it can lead to additional data errors due to movement or wait operations.  Alternatives to post-selection may achieve lower
overall failure rates due to avoiding such errors.
We present numerical simulation results comparing the logical error rates for the fault-tolerant [[7,1,3]] Steane code using techniques of ancilla verification vs. the newer method of ancilla decoding, as described in \cite{div07}.
We simulate QEC procedures in which rhe possibility of ancilla verification failures requires the creation and storage of additional ancillas and/or additional waiting of the data until a new ancilla can be created.
We find that the decoding method, which avoids verification failures, is advantageous in terms of overall error rate in various cases, even when measurement operations are no slower than others.
We analyze the effect of different classes of physical error on the relative performance of these two methods.
\end{abstract}


\maketitle 

\section{\label{sec:level1} Introduction}
\subsection{Ancilla post-selection in quantum error correction}
Quantum error-correcting codes (QECCs) provide a means to protect quantum data against noise by encoding quantum states into larger Hilbert spaces such that some class of error operations are correctable \cite{Shor95,steane96}. In a physically-realistic system, one must take into account that the correction operations themselves must be implemented using imperfect quantum operations.  In order for quantum error correction (QEC) operations to be effective in protecting the data, they
(and other operations on the data) must be implemented in a 
fault-tolerant way \cite{preskill98}, i.e., in a way such that a single faulty 
quantum gate cannot lead to multiple errors on the data. 

Some logical operations are naturally fault-tolerant, such as transversal gates in which every physical qubit is acted on by a separate gate,
and the absence of any ``cross-talk" between different physical qubits in an encoded block means there is no opportunity for errors to spread within that block.
However, QEC generally involves ancillary states which are used to carry away entropy and purge the data of errors.
These states must be very carefully prepared so that they do not spread errors to the data and this preparation must also operate fault-tolerantly.
Often, however, non-fault-tolerant circuits must be used for the initial ancilla preparation, and fault-tolerance is instead enforced by post-selection
(ancilla verification) 
in which only those ancillas which satisfy some measurement outcome after being created are subsequently used to interact with the data \cite{Shor95,steane96}.
Thus, it is not known beforehand whether a given ancilla will be used, and one may need to perform multiple ancilla creations before obtaining one which satisfies the post-selection criterion.  

This raises the question of how to ensure that a post-selected ancilla is available for QEC with sufficiently high probability so as not to significantly increase the overall failure rate of the QEC.
One may suppose that if an ancilla fails to pass verification, then the data would wait until a suitable ancilla is verified.
However, this is quite impractical since at any given point in the circuit, a large computation may have many parallel gating operations simultaneously performed.
If the data is required to wait, many qubits may have to wait, which could lead to many errors.
Furthermore, one must ideally have these ancillas created “as close as physically possible” to the data in order to avoid additional movement operations and associated errors.

Two obvious approaches are to either: 1) create a limited number of ancillas sequentially until post-selection is passed and to time the preparation 
so that it is likely that one will be available for use when needed, or 2) to create several ancillas in parallel so there is a high probability that at least one will pass \cite{Isailovic08}.
Both approaches have somewhat analogous disadvantages. In the first case, if an ancilla passes in the first try, sequential creation requires the ancilla to wait (and, in general, accumulate errors) until the ancilla will be used.  
Parallel creation avoids this, but has the problem that in a given physical architecture there will be a very limited number of ancillas that can be created close to the data.  
Thus, a verified ancilla may be created some distance away and need to be moved into contact with the data so the ancilla will accumulate errors from movement operations.
In either case, we must be prepared to skip QEC altogether rather than holding up the entire computation.  

In this paper we consider addressing this problem using {\it ancilla decoding} \cite{div07}, a technique originally devised to address the different problem of slow qubit measurements.
As we discuss below, ancilla decoding removes the need for post-selection and thus guarantees that any created ancilla can be used with the data.
In this case the ancilla can, in principle, be created in close proximity to the data and QEC performed without any additional movement or waiting and QEC need never be skipped.
Depending on the circuit layout and gate errors and timings, this may result
in lower overall logical failure rates and hence, in the case of a computation using a concatenated QECC, fewer resources required to achieve a given logical error rate.  
Since fault-tolerant methods and quantum error correction account for a relatively large 
amount of the resources used to perform reliable quantum computation, such savings 
are quite valuable.    In this work, we compare the performance (in terms of the logical error rate introduced in a noisy QEC procedure) of ancilla decoding with that of ancilla verification
procedures in order to demonstrate scenarios where decoding is advantageous even when measurement is no slower than any other operation.

We initially consider a naive ``series'' verification scenario, in which ancillas are created and verified until verification is passed (with the data
being held in (noisy) memory while additional verifications occur if the first is unsuccessful).  This avoids the situation where the QEC fails completely,
but, as discussed above, is unrealistic as part of a larger computation, since the duration of the QEC is unpredictable.

We additionally compare decoding to more realistic scenarios for both series and parallel ancilla verification.
In order to have a high probability for the ancilla to pass, two attempts are made to create the ancilla in either series or parallel.  (This is a physically motivated constraint given the discussion above of the sources of errors.)  
In the series case,  Figure~\ref{fig:series_model}, the ancilla created either passes on the first attempt or does not.  If it does, it is stored until it is needed.
If it does not, a second attempt is made.  If the second attempt also fails, we skip the QEC process altogether.  In the parallel case, Figure~\ref{fig:parallel_model}, two ancillas are prepared in parallel.
If the one closest to the data passes, it is used.  If it does not, the second one is tested.  If this ancilla passes, it is swapped (using a set of SWAP operations composed of CNOT gates) with the one that is closest.
If it also fails, the QEC process is skipped.  In all cases
we assume that the ancillary systems are created as close as possible to the data so that the series, first parallel,
and the decoding ancilla are all assumed equally close to the data in their respective implementations.  While not necessarily applicable to every physical layout,
this should be the case for many two-dimensional systems, giving some generality to our results while taking physical constraints into account.  We discuss layout further
in Section \ref{sec:layout}.

\subsection{The Steane code and ancilla creation}

In this section we compare the logical error rate $P_L$ obtained when performing fault-tolerant QEC operations for the well-known [[7,1,3]] Steane code \cite{steane96}, using either
ancilla verification or decoding.  The Steane code encodes a single logical qubit into the state of seven physical qubits, and can correct any error on a single physical qubit. 
It is a CSS stabilizer code \cite{Calderbank:96,Shor95,Steane97,Gottesman:97a,Gottesman:97b}, whose list of stabilizer generators is given in Table \ref{tab:stab} in the appendix.  In our analysis we consider the Steane ancilla technique for QEC, in which the code
stabilizers are measured by copying error information to ancillas in encoded logical states. 

\begin{figure*}
\includegraphics[scale=0.7]{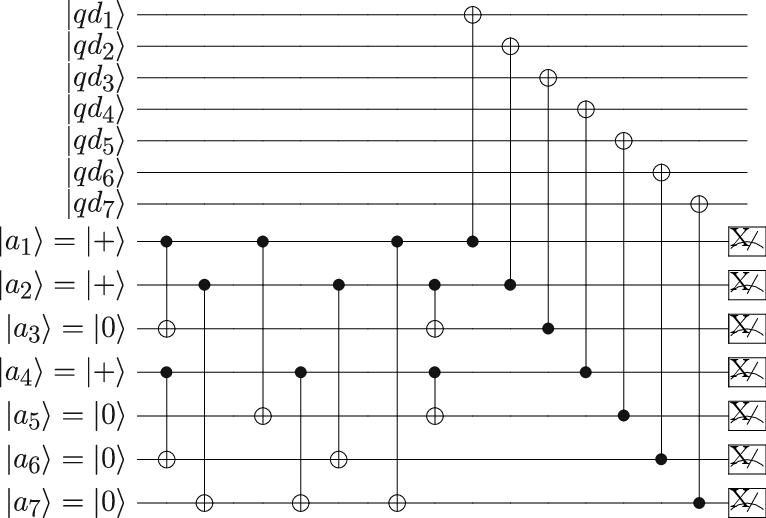}
\caption{\label{fig:zero} A Steane code circuit with non-fault-tolerant Steane syndrome extraction, where $\ket{qd_1}$, $\ket{qd_2}$ $...$,$\ket{qd_7}$ are the data qubits and $\ket{a_1}$, $\ket{a_2}$, $...$,$\ket{a_7}$ the ancilla qubits,
encoded via CNOT gates to produce the logical ancilla state (i.e., $\ket{0_L}$).
The ancilla-data interactions are implemented by a set of transversal CNOT gates, where any $Z$ error in the data will propagate to the ancilla. Ultimately, the ancilla is measured in the X-basis to determine the error syndrome.}
\end{figure*}

A (non-fault-tolerant) picture of part of the QEC process is shown in Figure \ref{fig:zero}, in which the data (qubits $\ket{qd}$)
interacts with a prepared logical $\ket{0_L}$ (where subscript $L$ denotes logical states) ancilla state (qubits $\ket{a}$)
to determine the error syndrome for $Z$ errors
(an analogous interaction occurs with a logical $\ket{+_L}$ to correct for $X$ errors).
We see, for example, that for fault-tolerance the $\ket{0_L}$ ancilla, which interacts with the data as the source for a transversal CNOT gate,
needs to be prepared so that a single gate failure will not cause the ancilla to have multiple Pauli $X$
errors (likewise with the $\ket{{+}_L}$ state and $Z$ errors).  Such  errors  would get transferred to the data, as can occur in the circuit shown in Figure \ref{fig:zero}, since the CNOT gates involved in preparing $\ket{0_L}$
can propagate a single gate error to multiple qubits within the ancilla block. 

\begin{figure}
\includegraphics[scale=0.8]{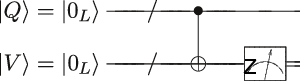}
\caption{\label{fig:verify} An example of a single ancilla preparation  and verification circuit taken from the Steane code with
Steane syndrome extraction. Where $\ket{Q}$ is the encoded ancilla state
and $\ket{V}$ is the encoded verifier state. Any $X$ errors which have occurred in the creation
of $\ket{Q}$ will propagate to the verifier via the transversal CNOT. }
\end{figure}

The standard
approach for ensuring this using ancilla verification (which occurs between ancilla preparation and data interaction) is shown in Figure \ref{fig:verify}.  Passing the verification procedure is dependent on the outcome of the transversal $Z$ measurement
performed on the verifier.  An analogous procedure occurs for $Z$ errors and the $\ket{+_L}$.

\begin{figure}
\includegraphics [scale=0.4]{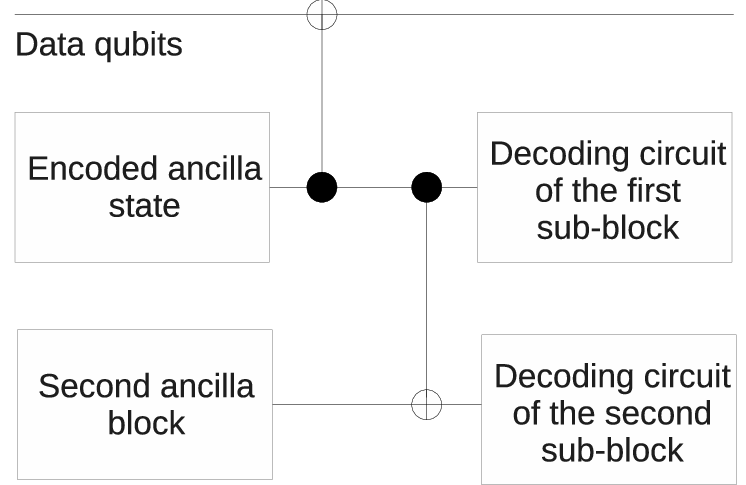}
\caption{\label{fig:decoder}The ancilla decoding procedure.  After interacting the data with a non-post-selected ancilla a decoding procedure is applied to determine the error
syndrome and any multi-qubit errors which may have been transferred to the data.  A second block allows one to distinguish between errors from encoding and decoding}
\end{figure}

The alternative method of ancilla decoding \cite{div07} is illustrated in Figure \ref{fig:decoder}. The basic principle is that after
interacting the ancilla with the data,  one applies a decoding operation to the ancilla to extract both the error syndrome (as usual) and
to determine the presence of any errors on the ancilla which may have been transferred to the data
($X$ or $Z$, depending on the ancilla in question).  In addition, the ancilla interacts with a second ancilla block which is in a product of $\ket{0}$ states (i.e. state $\ket{0}^{\otimes 7}$),  which is also decoded after interacting with the data.
This allows one to distinguish between errors of the first ancilla acquired during ancilla encoding
(which will have been propagated to the data and need correcting) and those during decoding (which will not be transferred to the
data or to the second block).  The decoding circuit is simply the time-reversal of the ancilla creation circuit as illustrated for state $\ket{0_L}$ in Figure \ref{fig:zero},
with the CNOT
gates applied in reverse order and creation of $\ket{+}/\ket{0}$ states replaced with measurement in the $X$/$Z$ bases respectively.

Any propagated errors (single or multi-qubit) can then be corrected on the data.  An important factor
in the success of this technique is that, for the purposes of fault-tolerance, only ``first-order'' patterns of errors, which can be
caused by a single gate failure (though they may still affect multiple ancilla qubits through error propagation, or through errors on two-qubit gates)
need to be corrected.  Furthermore, many error patterns are equivalent up to stabilizer operations.  This allows the limited
information from the decoder to be sufficient to correct any such errors.  Secondly, different classes of errors ($X$ and $Z$ or vice versa)
are respectively detected by the standard syndrome measurement and by the additional decoder syndrome.  For example, a $\ket{0_L}$ ancilla
is used to determine the syndrome for $Z$ errors on the data, but is at risk of propagating $X$ errors to the data, with the additional
decoding giving information about the latter.  The use of this method,  therefore,  means that any ancilla
which is created (regardless of first-order errors) can be used in QEC, removing the need for verification.
The encoding circuit is illustrated in Figure \ref{fig:CNOT} in the appendix,
each gate in that circuit being  faulty and capable of producing an error which will propagate (either as a single or multi-qubit error) to the data via the transversal CNOT gates
(as shown in Figure \ref{fig:zero}). For example, if a single fault has been produced by the two
 output channels  of  CNOT$-7$, or $-8$, or $-9$ (in Figure \ref{fig:CNOT}), then this error will propagate as a two-qubit error to the data.  This is discussed in more detail in the appendix.

As mentioned above, ancilla decoding was originally proposed to avoid long waits for ancilla verification in the case of
slow measurements.  Since, unless performing non-Clifford-group operations on our data, we can operate in the ``Pauli frame'' (merely classically recording the necessary corrections and
updating the stabilizer accordingly rather than actually applying them), there is no corresponding need for the data to wait
for the outcome of the ancilla decoding operations.  One could also partially avoid the problem of waiting for verification by simply 
beginning the verification well in advance, but even for fast measurements, this does not avoid any delays due to having to restart preparation if verification fails or errors due to waiting if the ancilla needs to be stored. 
In our analysis we consider the case of measurement operations no slower than any other gate, and show that
decoding still gives an advantage for this reason, even if verification failure is rare.

As discussed above, we are interested in the limitations imposed by practical QEC architectures where a reliable ancilla may not always be available when needed.  
We simulate cases where ancilla creation occurs serially or in parallel and a failed ancilla verification
requires the availability of a replacement, and compare the performance (in terms of the overall $P_L$) of a QEC 
procedure under these circumstances to QEC using the decoding method.  While we consider the  specific case of the Steane code using the Steane ancilla technique, it is conjectured 
that a similar decoding procedure can be found for other 
CSS QECCs \cite{div07}.  

\section{\label{sec:level2} Simulation procedure}
To compare methods of ancilla interaction, we performed Monte Carlo simulations of the complete QEC procedure (so interaction of the data with
two ancillas, one for the correction of each of $X$ and $Z$ errors) implemented using faulty gate operations.  Our simulation software was QASM-P,
software based on QASM by Cross \cite{cross06}.  Initially, all gates were simulated using the following common stochastic error model for depolarizing noise,
a function of a single error probability $p$:
\begin{enumerate}
\item 	Attempting to perform a data qubit identity I (wait operation), but instead performing a single qubit 
        operation $X$,  $Y$, or  $Z$, each occurring with probability $p/3$.
\item	Attempting to initialize a qubit to $\ket{0}$ or $\ket{+}$, but instead preparing $\ket{1}$ and $\ket{-}$, 
         respectively, with probability $p$.
\item Performing a $Z$-basis or $X$-basis qubit measurement, but reporting the wrong value with probability $p$.
\item  Attempting to perform a CNOT gate, but instead performing a CNOT followed by one of the two-qubit
            operations $I\otimes X$, $I\otimes  Y$, $I\otimes  Z$, $X \otimes I$, $X\otimes X$, $X \otimes  Y$, $X \otimes  Z$,  $Y \otimes I$,  $Y \otimes X$,  $Y\otimes  Y$,  $Y \otimes  Z$, $Z \otimes I$, $Z \otimes X$, $Z \otimes Y$,  or  $Z \otimes Z$, each with probability $p/15$.
\end{enumerate}
(As discussed later, when comparing more realistic series and parallel models we independently adjusted the error rates for the wait and CNOT operations).
We considered a range of values of $p$ below the threshold for Steane code of roughly $10^{-4}$ \cite{steane03}.
All gates are assumed to have the same duration, with wait operations implemented by applying the identity gate the appropriate number of times.

To determine $P_L$ for the QEC procedure, the data was first prepared, without error, in a logical eigenstate.  The QEC procedure was then
performed using the error model above.  Finally, the data was checked for logical errors.  To leading order (where the output data state
has no more than two physical qubit errors) we can treat logical $X$, $Y$ and $Z$ errors on the data as mutually exclusive,
thus in this case the overall logical error rate $P_L$ can be approximated as $P_X+P_Y+P_Z$ where, e.g., $P_X$ is the probability of the data receiving a logical $X$ error.
We determine $P_L$ by performing simulations using logical $X$, $Y$ and $Z$ eigenstates for the data, where, for example $E_X=P_X+P_Y$,
where $E_X$ is the simulation-determined logical error rate on eigenstates of the $Z$ basis.   From the Monte
Carlo simulation we therefore obtain $P_L$ as a function of the underlying physical error rates.  

\subsection{Simple series simulation}
As discussed above, in our simple model for a series procedure of indefinite duration, we assume the ancilla is initially prepared an appropriate length of time prior to the QEC so that the data is  not required to wait for ancilla preparation,
so long as the first ancilla successfully passes verification. However, if ancilla creation fails, the data is held
until a new ancilla can be created and verified.  Holding the data continues until verification is passed, at which point the QEC continues. 
We apply wait operations to the data for a total of $6$ time steps (the number required to create and verify a new ancilla) for every failed verification, with no upper bound to the number of failures that can occur. 
Once an ancilla has successfully passed verification the QEC procedure continues.
To determine the extent to which the verification process is affected by this data holding (as opposed to any other element),
we additionally simulate an unrealistically optimistic case of verification where no additional holding is required if verification fails.

Additionally, we generated simulation failure rates in which particular classes of errors were considered in isolation, i.e., with other errors turned off.
As in, e.g., \cite{fowler12}, we define these classes of errors as follows:
\begin{itemize}
\item {Class 0}: errors from preparation and measurement.
\item {Class 1}: errors from single-qubit gates (i.e., wait operations, since correction Pauli gates are error-free).
\item {Class 2}: errors from two-qubit gates (i.e., the CNOT gate).
\end{itemize}
We additionally performed simulations of more realistic scenarios (preserving data synchronicity) as described below.

\subsection{2-ancilla series simulation}
\begin{figure}
\includegraphics [scale=0.4]{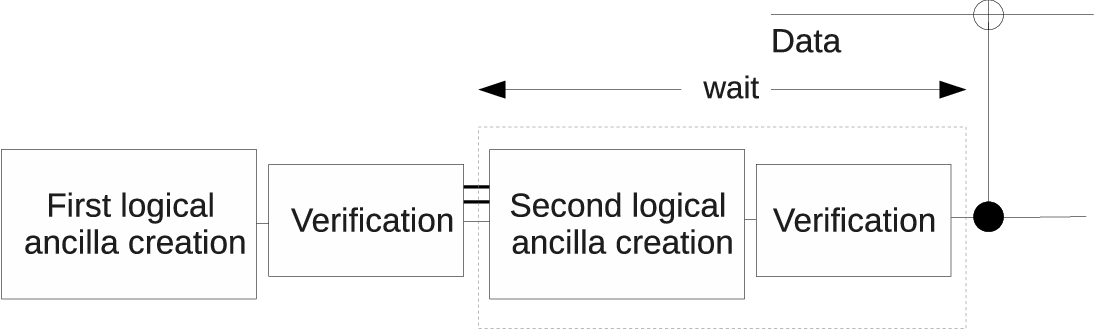}
\caption{\label{fig:series_model}Series ancilla verification. If the first logical ancilla state has passed the verification
test, then, it waits for $6$ time steps (duration of ancilla creation and verification) before interacting with the data.
Otherwise (if verification fails), we prepare 
and verify the second logical ancilla state and use that for QEC if verification passes.  Operations in the dashed box are conditioned
on the first verification, either they or the $WAIT$ operation takes place. If no verifications pass the QEC is abandoned. }
\end{figure}

In our 2-ancilla series simulation, an ancilla is initially prepared and verified ahead of the data's arrival.  If verification fails, a second ancilla is prepared and verified.
To preserve a consistent timing for the QEC operation, interaction of the ancilla with the data with the ancilla always occurs after this second verification would have occurred.
Thus, if the first ancilla passes verification it waits for a further 6 time steps before interacting with the data (which is assumed to arrive `` just in time'' and thus
never waits).  Hence, unlike the naive simulation, additional waiting occurs in the default scenario (i.e. where the first ancilla passes), and is applied to the ancilla rather
than the data. See $\bf{ Figure}$  $\ref{fig:series_model}$.

\subsection{2-ancilla parallel simulation}
\begin{figure}
\includegraphics [scale=0.4]{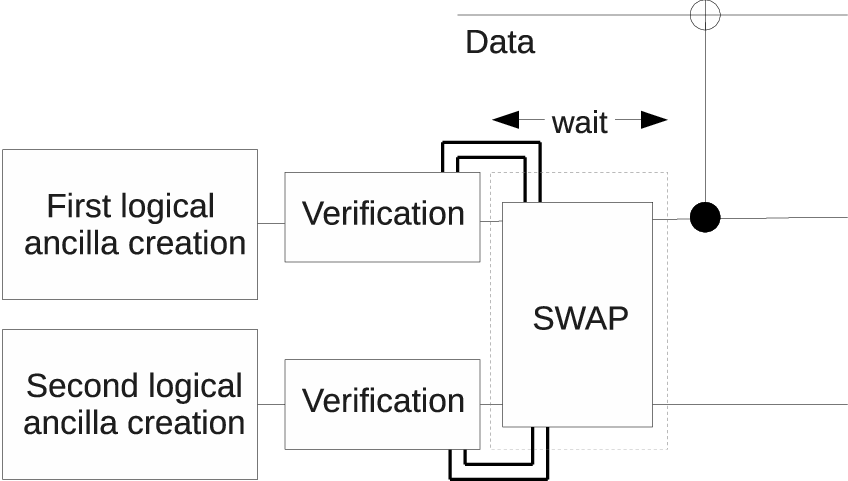}
\caption{\label{fig:parallel_model} Parallel ancilla verification. If the first logical ancilla state has passed the verification
test, it waits for $3$ time steps (the duration of a $SWAP$ gate) before interacting with the data
(example shown is for correction of $Z$ errors).
Otherwise (if only the second ancilla passes verification), a $SWAP$ operation
is applied between two states, and the verified ancilla used for QEC.  Operations in the dashed box are conditioned
on the first verification, either they or the $WAIT$ operation takes place.  If no verifications pass the QEC is abandoned.}
\end{figure}

In the 2-ancilla parallel simulation, both ancillas are prepared and verified simultaneously.  One ancilla (the first) is assumed to be ``adjacent'' to the data, in the sense that it can
interact with the data qubits without any additional movement operations.  The second ancilla is considered adjacent to the first ancilla (but not the data).  If verification of the first ancilla
fails but the second passes, a transversal SWAP operation (consisting of 3 CNOT gates in sequence, the first and third using one qubit as the control and the second using the other qubit)
is performed between the two ancillas so that the verified ancilla is now adjacent to the data, after which the QEC proceeds.
As before, for timing consistency, the data is assumed to arrive just in time for interaction after this SWAP has occurred.  Thus, if the first ancilla passes verification,
it undergoes an additional wait operation (the duration of the SWAP gate)  for 3 time steps   before interacting with the data. See Figure \ref{fig:parallel_model}.

\subsection{Layout considerations in simulation}\label{sec:layout}
\begin{figure}
\includegraphics [scale=0.5]{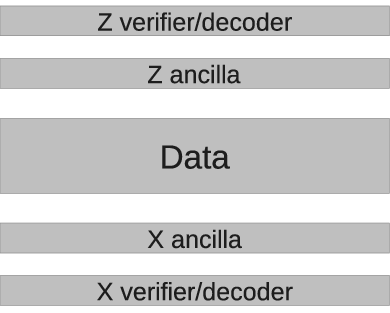}
\caption{\label{fig:layout}Underlying layout for simulations.  Ancillas are created in separate interaction regions adjacent both the the data and additional verification or
decoding ancillas.}
\end{figure}

We do not model the qubit layout in detail: our simulations (both simple and more realistic) for series verification and decoding do not feature any movement
operations.  However to roughly simulate the additional errors brought about by ancilla movement in the parallel case, we use the model illustrated in Figure \ref{fig:layout}.
Qubits for each logical block are positioned in ``interaction regions'' (the horizontal rectangles).  Within the same region any two qubits may interact without any additional movement.
However, for two logical blocks to transversally interact, they must be in adjacent regions (we model the interaction regions as having limited capacity - otherwise we could simply
place every logical block in the same region - in this case, equal to one logical block (7 qubits)).
Thus the series and decoding approaches do not require any additional movement operations; as illustrated, we may prepare the ``primary'' ancilla (which interacts directly with the data)
in the region adjacent to the data, and the verifier ancilla (or additional ancilla for decoding), which only needs to interact with the primary ancilla, in the region on the far side
of the primary ancilla.  We do this on one side of the data for the QEC used to correct $X$ errors, and on the other for the QEC for $Z$ errors, since we need to begin
preparation of the ancilla for one QEC while the other is still taking place, so the data requires no additional waiting between these operations.
\subsubsection{Layout in parallel verification}\label{sec:parlayout}
\begin{figure}
\includegraphics [scale=0.5]{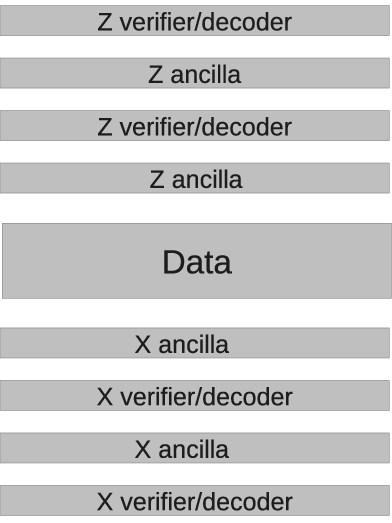}
\caption{\label{fig:layout-parallel}Layout for parallel ancilla verification.  If ancilla 1 fails verification, ancilla 2 must perform two SWAP operations to reach the data.}
\end{figure}

In the parallel verification case, if we are permitted one logical block per region, we arrange the ancillas as illustrated in Figure \ref{fig:layout-parallel}.
Thus secondary ancillas must be moved to interact with the data, giving additional errors.  We implement ``movement'' of this kind by applying SWAP gates (consisting of 3 CNOT gates)
between adjacent blocks.  Thus, to ``move'' a secondary ancilla adjacent to the data we must apply two SWAP gates to move it past both the original ancilla and the original
verifier.
However this means in our model that parallel verification can easily be seen to be worse than the series case, since the two SWAP gates consist of 6 total operations,
or the same amount of time as required to create and verify a new ancilla in the series case.  Thus the time required post-failure to have a new ancilla available is no less,
and additional errors are incurred from the SWAP gates, so the series method is clearly better in terms of errors incurred.  Since we are interested in demonstrating where decoding improves
on the (best available) verification methods even for fast measurements, we note this result and do not perform explicit simulations for this parallel case.

This result is, of course, dependent on our layout choices and our chosen method for moving the data and second ancilla adjacent.
However, it illustrates the cost of parallel methods for ancilla creation - in this case, although not necessarily always, so high as to make series verification uniformly better - 
of having to move successfully verified ancillas into place.
For the series and decoding scenarios which we do simulate, the difference between techniques consists primarily of additional wait and CNOT operations, the former in additional
ancilla waiting for synchronicity, the latter in decoding operations, and relative performance should therefore depend 
primarily on the error rates of these operations.  We therefore performed simulations for both scenarios and for ancilla decoding over CNOT
and wait error rates ranging from $10^{-5}$ to $3\times 10^{-4}$,
with all other gates fixed at an error rate of $10^{-5}$.

There is also, in the series case, the possibility that both ancillas will fail verification and no QEC occurs (which never occurs with decoding).
This is obviously a bad outcome (since, in a larger
computation, any prior error on the data would not be corrected).  However, it does not directly correspond to a logical error induced by the QEC.
Since we are interested in demonstrating that decoding
can be advantageous even when measurement is not slow, we simply rerun any such scenarios, which do not contribute to final pass/fail statistics.
Qualitatively, then, our use of $P_L$ slightly overestimates the performance of the verification techniques in this respect.

\section{\label{sec:level3} Results and analysis}
\subsection{Simple series simulation}
\begin{figure}
\includegraphics [scale=0.3]{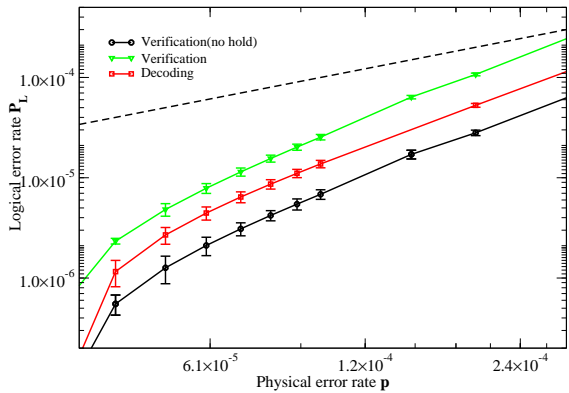}
\caption{\label{fig:compare}(Color online) Logical error rate as a function of physical error rate for three QEC techniques.  Dashed line represents $p=P_L$.}
\end{figure}

A direct comparison of the $P_L$ for all three QEC techniques (decoding, verification, and a naive verification without additional waiting) is shown for the simple series simulation in Figure \ref{fig:compare}.
The results show an advantage for the decoding technique over verification over the whole range of errors considered, larger at larger error rates with a roughly $2$-fold reduction as the maximum.  Comparison with the $P_L$ values for verification without additional waits, which are lower still, indicates as expected that the increased errors in the verification technique are due to ancilla failures, and that in the absence of such failures the additional operations required for ancilla decoding lead to a slightly higher error rate.

\begin{figure}
\includegraphics [scale=0.3]{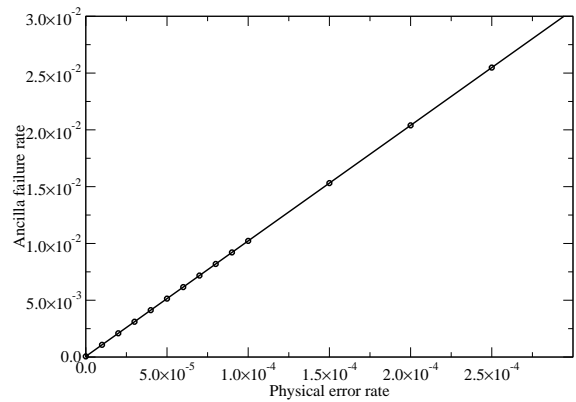}
\caption{\label{fig:ancillafail}The ancilla failure rate vs. physical error rate for the [[7,1,3]] Steane code with Steane ancilla}
\end{figure}

The rate of ancilla failure as a function of $p$ is plotted in Figure \ref{fig:ancillafail}.
Since ancilla verification detects single errors on the ancilla, this is roughly equal to (number of error locations in ancilla creation circuit)$\times p$, and the rate scales linearly with $p$ as expected. Thus at low physical error rates ancilla failure occurs in a proportionally low fraction of QEC events, but still contributes significantly to the logical error rate.

\begin{figure*}
\includegraphics [scale=0.6]{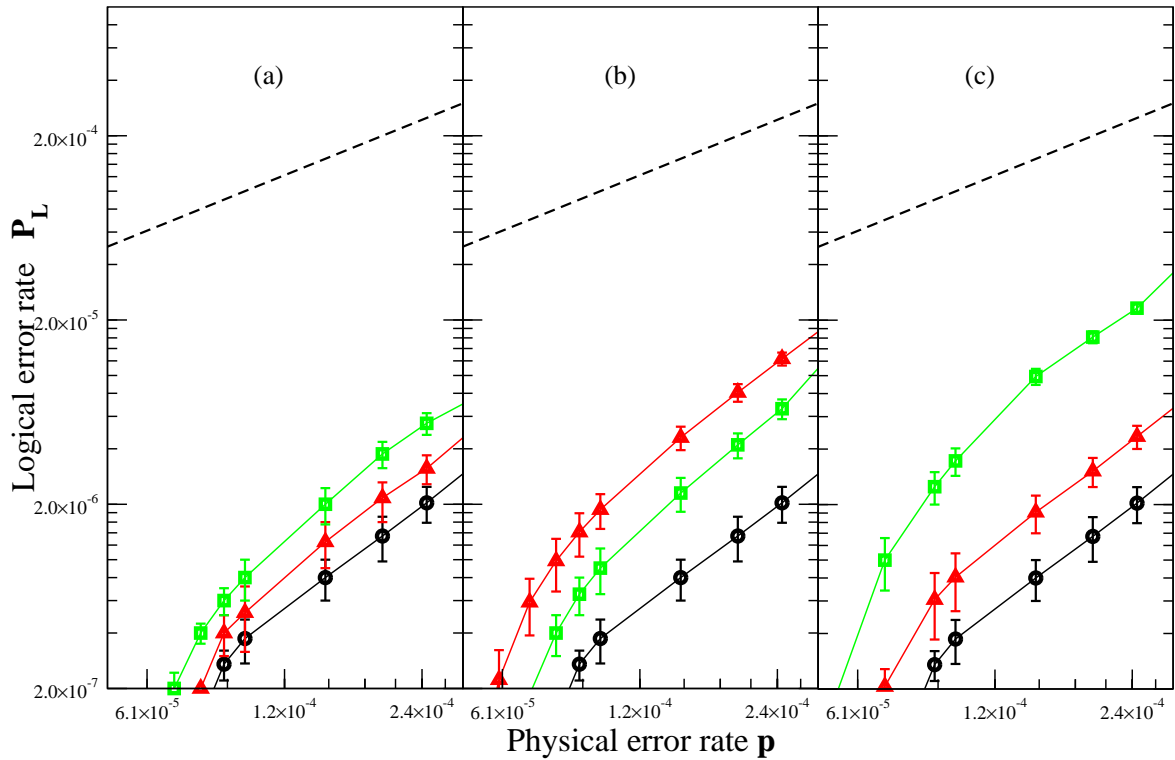}
\caption{\label{fig:class}(Color online) Logical error rate $P_L$  vs. 
physical error rate $p$ for class-0 errors (black circle points), class-1 errors (red triangle points),
 and class-2 (green square points) for (a) verification (without additional waits), (b) verification and (c) decoding. Dashed line represents $p=P_L$.}
\end{figure*}

Comparing the error rates by class in Figure \ref{fig:class} we see that the largest source of error in verification is due to wait operations, while the other two techniques are dominated by CNOT errors.
This emphasizes the need to take into account non-deterministic operations (such as verification) when assessing the impact of gate errors on QEC procedures.

\subsection{2-ancilla series simulation}
\begin{figure*}
\includegraphics [scale=1]{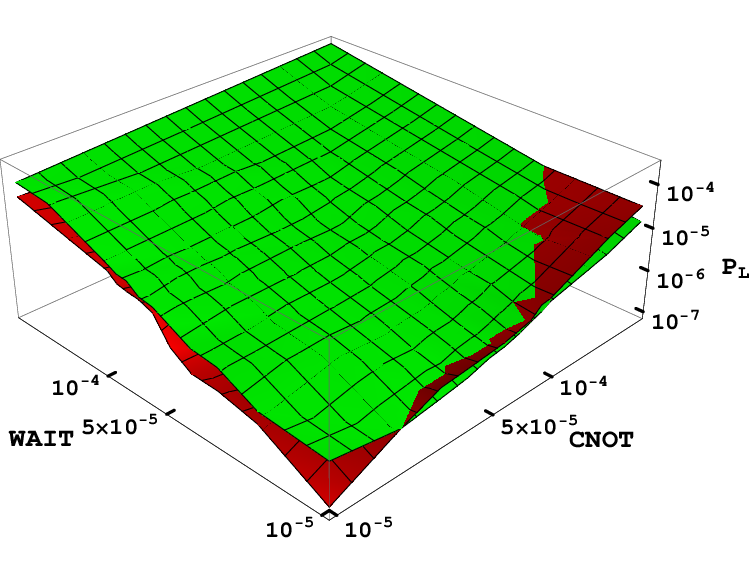}
\caption{\label{fig:series}(Color online) Logarithmic plot of logical error rate $P_L$ vs. physical error rate of the $CNOT$ and $WAIT$ gates 
for the decoding technique and verification technique with the two logical ancillas in series. The red sheet represents the decoding technique, 
and the green sheet represents the verification technique.}
\end{figure*}

Figure \ref{fig:series} shows logical failure rates for the series verification technique and the decoding technique,
over a range of CNOT and wait error rates.  In qualitative terms, these generally are as expected, but importantly
show when decoding is the better technique. 
We see again that decoding outperforms verification (i.e. has a lower logical failure rate) for a large range of error rates.  In contrast
with our simple series simulation where all gates had the same error rate (and decoding was consistently superior), we see, however, that verification outperforms decoding
in the range of low CNOT error and large wait error.  This is as expected, given that, relative to each other, verification requires additional wait operations
(to recreate ancillas) and decoding additional CNOT operations (to perform decoding).  As discussed in section \ref{sec:parlayout},
the parallel approach will be uniformly worse than the series approach in this model, so where decoding outperforms series verification it will also outperform
parallel verification.

\section{\label{sec:level5} Conclusions}
Using Monte Carlo simulations, we have compared the logical error rate, $P_L$, for implementations of the Steane code with Steane ancillas using different QEC techniques:
ancilla verification with serial and parallel production of multiple ancillas, and the ancilla decoding procedure.  
We find that, even when measurement times are no longer than those for other operations and verification
failures are rare, the decoding procedure, though otherwise more complex, is often advantageous in avoiding additional data waiting and/or movement due to verification failures
and leads to lower failure rates.

As the data shows, the relative performance of the various techniques is highly dependent on the underlying error rates of the individual operations.  A model where gates
have equal errors, ancilla creation continues serially until success and all logicl blocks are assumed adjacent to those with which they interact consistently produces higher logical error rates
for verification than for decoding, but, for example, small wait errors and large CNOT errors can result in better performance for verification. 
While we have endeavored to make plausible assumptions with regard to our simulations, there is of course much scope for more fully mapping
out the performance dependence on different techniques depending on individual gate errors and durations, and physical layout.  The latter,
in particular will likely make a great difference to the relative performance of the parallel ancilla technique.  For example, the superiority
of series-based verification in our case was a direct result of ancilla ``movement'' (via SWAP operations) being no faster than ancilla
recreation.  A model with faster ancilla movement or longer ancilla encoding time could easily change this.  We note that an explicit analysis of the effect of decoding
in a proposed layout for trapped-ion-based quantum computing has recently been published \cite{Tomita13}.  

 Our emphasis is
on the fact that decoding shows demonstrably better performance in a range of cases, even when measurement is not slow, and that this advantage becomes apparent
when one considers dealing with the practical consequences of verification failure: having to either recreate or move into position additional ancillas.
Moreover we note that the additional errors induced by the latter can make a significant
difference to overall error rates, and should be considered when assessing the performance of QEC techniques, including (as our work does not currently do)
the consequences in a many-stage computation of, on occasion, occasionally failing to verify any ancillas, and the optimal choice of techniques in this scenario.

\begin{acknowledgments}
We thank Panos Aliferis and Russell Ceballos for useful discussions and an anonymous referee for their helpful comments.  MSB thanks Jason Twamley and the Center for Engineer Quantum Systems where part of this work was completed. 
Supported by the Intelligence Advanced Research Projects Activity (IARPA) via Department of Interior National Business Center contract number D12PC00527. The U.S. Government is authorized to reproduce and distribute reprints for Governmental purposes notwithstanding any copyright annotation thereon. Disclaimer: The views and conclusions contained herein are those of the authors and should not be interpreted as necessarily representing the official policies or endorsements, either expressed or implied, of IARPA, DoI/NBC, or the U.S. Government.   
\end{acknowledgments}

\appendix*
\section{\label{sec:errortrack}Details of ancilla decoding}
In this section we describe the possible ancilla errors due to a single fault in the encoding circuit, and the corresponding error syndromes when using the ancilla decoding technique. This is an expansion of the summary given in \cite{div07}.  We will consider the case of $X$ errors due to faults  in the encoding circuit for $\ket{0_L}$; the case of $Z$ errors when encoding $\ket{+_L}$ is directly analogous. 

\begin{figure}
\includegraphics*[scale=0.8]{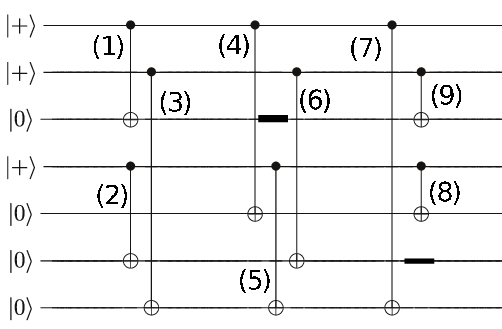}
\caption{\label{fig:CNOT} Encoding circuit for $\ket{0_L}$}
\end{figure}

Figure \ref{fig:CNOT} illustrates the circuit for encoding $\ket{0_L}$, which possesses two types of gates: (a) single qubit gates defined as preparation of $\ket{0}$ and $\ket{+}$ with waiting gates represented by bold line segments in Figure \ref{fig:CNOT}, as well as (b) two qubit gates defined as CNOT gates.  The decoding circuit can uniquely identify any single-qubit error.  Two qubit-errors caused by a single fault will either be errors on both outputs of a CNOT gate, or a single error which at some point propagates to two errors via CNOT, thus it suffices to further consider only the case
of a CNOT gate on which both outputs have $X$ errors.  These are listed in Table \ref{tab:CNOT} according to the encoding circuit labeling given in Figure \ref{fig:CNOT}.

As shown, dual output $X$ errors propagate to single $X$ errors to the data in the cases of CNOT gates 4, or 5, or 6; or as two $X$ errors to the data in the cases of CNOT gates 7, or 8, or 9.  See Table \ref{tab:CNOT}.  However, for CNOT gates 1, or 2, or 3, dual $X$ errors in the two output channels will propagate as four $X$ errors on the data.  These are equivalent to the $X$-stabilizer generators of the Steane Code, which accordingly do not affect the data.  The stabilizer generators of the Steane Code are listed in \ref{tab:stab}. Finally, for CNOT gates 4, or 5, or 6; the single faults which are produced in the two output channels will propagate as a dual $X$ errors on the data.

\begin{table}
\caption{Examples of output errors produced by $X$ errors on both outputs of CNOT gates in the $\ket{0_L}$ encoding circuit (Figure \ref{fig:CNOT}).}
\centering
\begin{tabular}{c c }
\hline\hline
    CNOT& Multi-qubit data errors \\
\hline
1&$X_{1} X_{3}X_{5}X_{7}\equiv G_{1}$ \\
2&$X_{4} X_{5}X_{6}X_{7}\equiv G_{2}$\\
3&$X_{2} X_{3}X_{6}X_{7}\equiv G_{3}$\\
\hline
4&$X_{1} X_{5}X_{7}\equiv G_{1}X_3$\\
5&$X_{4} X_{5}X_{7}\equiv G_{2}X_6$\\
6&$X_{2} X_{3}X_{6}\equiv G_{3}X_7$\\
\hline
7&$X_{2} X_{3}\equiv G_{3}X_6X_7$\\
8&$X_{4} X_{5}\equiv G_{2}X_6X_7$\\
9&$X_{1}X_{7}\equiv G_{1}X_3X_5$\\
\hline\hline
\hline
\end{tabular}
\label{tab:CNOT}
\end{table}

\begin{table}
\caption{The stabilizer generators of the Steane code.}
\centering
\begin{tabular}{c c}
\hline\hline
X-stabilizer& Z-stabilizer \\[0.5ex]
\hline
$G_{1}= X_{1} X_{3}X{_5}X_{7}$&$G_{1}= Z_{1} Z_{3}Z_{5}Z_{7}$\\
$G_{2}= X_{4} X_{5}X_{6}X_{7}$&$G_{2}= Z_{4} Z_{5}Z_{6}Z_{7}$\\
$G_{3}= X_{2} X_{3}X_{6}X_{7}$& $G_{3}= Z_{2} Z_{3}Z_{6}Z_{7}$ \\               
\hline\hline							
\hline
\end{tabular}
\label{tab:stab}
\end{table}
These $X$ errors (single and double $X$ errors) on the data can be identified by the decoding circuit and so corrected.  As seen, every multi-qubit error, is, up to stabilizers, equivalent either to a single-qubit error or to one of three two-qubit errors,
\FloatBarrier
\bibliographystyle{apsrev4-1}
\bibliography{bib}
\end{document}